\documentstyle[12pt,fleqn,epsfig]{article}
\begin{document}


\begin{center}
{\Large \bf $\Omega$, $J/\psi$ and $\psi'$ Production in Nuclear
Collisions }
\end{center}
\begin{center}
{\Large \bf and Quark Gluon Plasma Hadronization}
\end{center}

\vspace{0.5cm}
\begin{center}
{\bf
M. I. Gorenstein $^{a,b}$, K. A. Bugaev $^{a,c}$} and {\bf
M. Ga\'zdzicki $^d$
}
\end{center}

\vspace{0.3cm}

$^a$ Bogolyubov Institute
for Theoretical Physics,
Kiev, Ukraine

$^b$ Institut f\"ur Theoretische Physik, Universit\"at  Frankfurt,
Germany

$^c$ Gesellschaft f\"ur Schwerionenforschung (GSI), Darmstadt, Germany

$^d$ Institut f\"ur  Kernphysik, Universit\"at  Frankfurt,
Germany

\vspace{1.0cm}

\begin{abstract}
\noindent The transverse mass spectra of $\Omega$, $J/\psi$ and
$\psi'$ in Pb+Pb collisions at 158~A$\cdot$GeV are studied within a
hydrodynamical model of the quark gluon plasma expansion and
hadronization. The model reproduces the existing data with the common
hadronization parameters: temperature $T=T_H \cong 170$~MeV and average
collective transverse velocity
 $\overline{v}_T\cong 0.2$.

\end{abstract}

 The possibility of observing the quark gluon plasma (QGP) in
nucleus-nucleus (A+A) collisions motivated an extensive experimental
program. Rich experimental results on hadron spectra and multiplicities
are now available for a broad range of collision energy and for various
colliding systems \cite{qm2001}. The anomalies observed \cite{sqm2001} in
the energy dependence of pion and strangeness production indicate
\cite{gago} that in central collisions of heavy nuclei the threshold for
QGP creation during the early stage of the reaction is in the region of low
CERN SPS energies ($\approx 40$ A$\cdot$GeV).

The equilibrium hadron gas (HG) model describes remarkably well the hadron
multiplicities measured in A+A collisions at top SPS \cite{HG}
and RHIC \cite{HG1} energies, where the creation of QGP is expected. The
extracted hadronization temperature parameter is 
similar for both energies $T_H=170\pm
10$~MeV. This is close to an estimate of the temperature $T_C$ for
the
QGP--HG transition obtained in Lattice QCD simulations at zero baryonic
density (see e.g. \cite{Karsch}). One may therefore argue 
that the QGP created
in high energy heavy ion collisions  hadronizes
into an (approximately) locally equilibrated HG and the chemical
composition of this HG is weakly affected by rescattering during the
expansion of the hadronic system \cite{stock}.

In our previous paper \cite{BGG} we formulated the hypothesis that the
kinetic freeze--out of $J/\psi$ and $\psi'$ mesons takes place directly at
hadronization and that those mesons therefore carry information on the
flow velocity of strongly interacting matter just after the transition to
the HG. Based on the measured $J/\psi$ and $\psi'$ spectra in Pb+Pb
collisions at 158 A$\cdot$GeV \cite{Jpsi} and using the hypothesis of
the statistical production of charmonia at hadronization
\cite{Ga1,Br1,Go:00} we extracted a mean transverse collective flow
velocity of hadronizing matter: $\overline{v}_T\cong 0.2$.

The effect of the rescattering in the hadronic phase was recently studied
within a ``hydro + cascade'' approach \cite{BD,Sh}. A+A collisions are
considered there to proceed in three stages: hydrodynamic QGP expansion
(``hydro''), transition from QGP to HG and the stage of hadronic
rescattering and resonance decays (``cascade''). The change from ``hydro'' to
``cascade'' modelling takes place at $T=T_{C}$, where the spectrum of hadrons
leaving the surface of the QGP--HG transition is taken as input for the
subsequent cascade calculations. The results of Refs.~\cite{BD,Sh}
suggest
that the transverse momentum 
($p_T$) spectrum of $\Omega$ may be weakly affected
 during the
cascade stage even for central Pb+Pb collisions at the top SPS energy.
This is because of the small hadronic cross section and large mass
of the $\Omega$
hyperon
\cite{Sorge}.
The corresponding calculations for charmonia are  not yet performed
within this model.

Thus we are faced with an intriguing problem: if the above considerations
for charmonia \cite{BGG} and $\Omega$ ~\cite{BD,Sh} are correct, their
$p_T$ spectra should be simultaneously reproduced using the same
hydrodynamic parameters, $T_H$ and $\overline{v}_T$. In this letter we
demonstrate that such a description is indeed possible. The transverse
mass $m_T$ ($m_T = \sqrt{p_T^2 + m^2}$, where $m$ is a particle rest mass)
spectra around midrapidity in Pb+Pb at the SPS (158~A$\cdot$GeV) were
recently measured for $\Omega$
by WA97 Collaboration \cite{Omega, Omega1} and for $J/\psi$
and $\psi'$ by NA50 Collaboration \cite{Jpsi}. These spectra will be the
subject of the present analysis.

\vspace{0.3cm} Assuming kinetic freeze--out of matter
 at constant temperature $T$,
the transverse mass spectrum of $i$-th hadron species (with mass $m_{i}$)
in cylindrically symmetric and longitudinally boost invariant fluid
expansion can be approximated as \cite{Heinz}:
\begin{equation}\label{hydro}
\frac{dN_i}{m_T dm_T}~
\propto~
m_T~ \int_{0}^{R}r dr~ K_1\left({ \frac{m_T \cosh y_T}{T} }\right)~
I_0\left({ \frac{p_T\sinh y_T}{T}}\right)~,
\end{equation}
where $y_T=\tanh^{-1}v_T$ is the transverse fluid rapidity, $R$ is the
transverse system size, $K_1$ and $I_0$ are the modified Bessel functions.
The spectrum (\ref{hydro}) is obtained under assumption that the
freeze-out occurs at constant longitudinal proper time $\tau =\sqrt{t^2-z^2}$,
where $t$ is the time and $z$ is the longitudinal coordinate. 
Thus the
freeze--out time $t$ is independent of the transverse coordinate $r$
\cite{Heinz}. 
The analysis of the numerical calculations of Ref.~\cite{Sh} 
shows that the latter is approximately fulfilled. 
The quality of the approximation made gets better for considered here
heavy particles and small transverse flow velocities because a possible
deviation from Eq.~(\ref{hydro})
is proportional to $p_T^2 v_T / (2 m_T T)$ 
and thus it decreases with increasing particle mass at
constant $p_T$.

In order to calculate (\ref{hydro}) the function
$y_T(r)$ has to be given. 
A linear flow profile, $y_T(r)=y_T^{max}\cdot r/R$, is often
assumed in phenomenological fits \cite{Heinz}.
The numerical calculations of Ref.~\cite{Sh} justify this assumption. For
heavy hadrons analysed in this work
the condition $m_i>>T$ is always satisfied and, therefore, the asymptotic
form for large arguments of $K_1(x)\sim x^{-1/2}\exp(-x)$ can be used in
Eq.~(\ref{hydro}). At SPS energies typical values of $v_T$ are small
($v_T^2<<1$) and consequently $\cosh y_T\cong 1+ \frac{1}{2}v_T^2$~ and
$\sinh y_T\cong v_T$. 

The experimental $m_T$--spectra
are usually parametrised by a function\footnote{
It corresponds formally to neglecting the transverse flow
($v_T\equiv 0$ in Eq.~(\ref{hydro})), but introducing instead an
``effective'' temperature.}:
\begin{equation}\label{T*}
\frac{dN_i}{m_T dm_T}~
\propto ~
\sqrt{m_T}~
\exp\left({ -~\frac{m_T}{T_i^*} }\right)~,
\end{equation}
where the inverse slope $T^*_i$ is extracted from the fit to the data.
However, when Eq.~(\ref{T*}) is considered as an
approximation of Eq.~(\ref{hydro}), the inverse slopes $T_i^*$ 
should depend on both $m_i$ and $p_T$.
The limiting cases of $T_i^*$ behaviour at low and high $p_T$ can be easily
studied using the small and large argument asymptotic of the modified
Bessel function $I_0$ in 
Eq.~(\ref{hydro}):
\begin{equation}\label{T1}
T^*_i(p_T\rightarrow 0)~=~\frac{T}{1~-~\frac{1}{2}~\overline{v}_T^2~
(m_i/T~-~1)}~\approx~ T~+ \frac{1}{2}~m_i~\overline{v}_T^2~,
\end{equation}
\begin{equation}\label{T2}
T^*_i(p_T\rightarrow \infty)~\equiv ~ T^*~=~\frac{T}{1~-~v_T^{max}~+~
\frac{1}{2}~(v_T^{max})^2}~,
\end{equation}
where the average velocity $\overline{v}_T$ in Eq.~(\ref{T1}) is defined as
$\overline{v}_T^2=\int_0^R rdr v_T^2(r)/\int_0^R rdr$. 
The maximum velocity $v_T^{max}$ in Eq.~(\ref{T2}) is related to
$\overline{v}_T$ as $\overline{v}_T^2=(v_T^{max})^2/2$
provided a linear flow profile is assumed, $v_T(r)= v_T^{max}\cdot r/R$.
Note that $T^*$ in Eq.~(\ref{T2}) is equivalent to the well known ``blue
shifted'' temperature, $T[(1+v_T^{max})/(1-v_T^{max})]^{1/2}$
(see e.g. \cite{Sh,Heinz}), calculated for $(v_T^{max})^2<<1$. The shape
of the ``high--$p_T$'' tail ($p_T>>m_i$) of the $m_T$ distribution
(\ref{hydro}) is ``universal'', i.e. $T^*$ given by Eq.~(\ref{T2}) is
independent of particle mass $m_i$. On the other hand, the inverse slopes
$T^*_i$ (\ref{T1}) at ``low--$p_T$'' are strongly dependent on $m_i$. Two
remarks are appropriate here. First, for heavy particles like $\Omega$ and
$J/\psi$ the term $\frac{1}{2}m_i\overline{v}_T^2/T$ 
in Eq.~(\ref{T1}) is not small compared
to one thus the second (approximate) equality in this equation is
violated. Second, a condition of the validity of Eq.~(\ref{T1}),
$p_T\overline{v}_T << T$, is too restrictive for heavy hadrons, e.g. for
$T\cong 170$~MeV and $\overline{v}_T\cong 0.2$ discussed below it leads
to $m_T-m_i << 0.3/m_i$~GeV$^2$/$c^{4}$. This means that Eq.~(\ref{T1}) is
valid for the values of $m_T-m_i$ which are much smaller than
0.2~GeV/c$^2$ for $\Omega$ and than 0.1~GeV/c$^2$ for $J/\psi$.

Summarising: (a) none of the asymptotic regimes ({\ref{T1}) or
(\ref{T2})
can be clearly seen in the experimental $m_T$ spectra, i.e. neither
``low--$p_T$'' ($m_T-m_i << 0.3/m_i$~GeV$^2$/$c^{4}$) nor ``high--$p_T$''
($m_T - m_i>>m_i$) approximations are useful ones (at least for studying
the available $m_T$ spectra of $\Omega$ and charmonia); (b) fitting the
experimental $m_T$ spectrum of $i$--th hadron species by Eq.~(\ref{T*}) one
finds in fact the ``average inverse slopes'' which depend not only on
particle mass $m_i$, but also on the $m_T-m_i$ interval covered in a given
experiment (see also Ref.~\cite{Sh}, where $T^*_i$ have been discussed
separately for $m_T-m_i< 0.6~$GeV/c$^2$ and for 0.6~GeV/c$^2 < m_T-m_i<
1.6~$GeV/c$^2$). 

For small values of $v_T$ relevant for our discussion
a  good approximation of Eq.~(\ref{hydro}) at $m_T-m_i<m_i$ can be
obtained by substituting the $v_T$ distribution in Eq.~(\ref{hydro})
by its average value $\overline{v}_T$ and by using large argument
$K_1$ asymptotic:
\begin{equation}\label{hydro1}
\frac{dN_i}{m_T dm_T}~
\propto ~
\sqrt{m_T}~
\exp\left({ -~\frac{m_T (1+\frac{1}{2}\overline{v}_T^2)}{T} }\right)
~I_0\left({ \frac{p_T\overline{v}_T}{T} }\right)~.
\end{equation}
We checked numerically that the values of the parameter
$\overline{v}_T$ extracted from the fits (see below) to Eqs.~(\ref{hydro})
and  ~(\ref{hydro1}) assuming a linear velocity profile
differ by about 5\%.

\vspace{0.3cm}
We turn now to the test of our hypothesis of the kinetic
freeze--out of $J/\psi$, $\psi'$ and $\Omega$ occurring directly at hadronization
i.e. at $T = T_H = 170$ MeV.
The $m_T$--spectra of these hadrons are measured around midrapidity 
\cite{Jpsi,Omega} for Pb+Pb collisions at 158 A$\cdot$GeV.
The fit to these data performed using Eq.~(\ref{hydro1}) with
$T = T_H = 170$ MeV yields $\overline{v}_T=0.194 \pm 0.017 $ and
$\chi^2/dof = 1.3$.
The value of $\overline{v}_T$ varies by $\mp 0.016$ when $T_H$ is changed
within its uncertainty $\pm 10$ MeV.
Note that $T$ and $\overline{v}_T$ parameters are anti--correlated.
A surprisingly good agreement (see Fig.~1) of our model with the data on 
$m_T$--spectra serves as a strong support of the hypothesis of
statistical nature of $J/\psi$ and $\psi'$ production
\cite{Ga1} and their kinetic freeze--out occurring directly at hadronization
\cite{BGG}.

The dependence of the $J/\psi$ and $\psi'$ transverse mass spectrum on the
centrality (quantified by the neutral transverse energy $E_{T}$) of Pb+Pb
collisions at 158~A$\cdot$GeV was also measured by NA50 Collaboration
\cite{Jpsi}. An increase of $\langle p_{T} \rangle$ and $\langle p_{T}^2
\rangle$ from peripheral collisions to the most central collisions can be
explained, within our approach, by an increase of the model parameter
$\overline{v}_{T}$ with $E_{T}$. Note that the increase of mean flow
velocity and consequently an increase of $\langle p_{T} \rangle$ and
$\langle p_{T}^{2}\rangle$ with a centrality of the collision as well as
an increase of $\langle p_{T}\rangle$ and $\langle p_{T}^{2}
\rangle $ with particle mass $m_{i}$ are characteristic features of
hydrodynamics. In contrast to $J/\psi$ and $\psi^{\prime}$ mesons the
$m_{T}$--spectra of the Drell--Yan pairs (dileptons with invariant mass
$M>4.2$~GeV/c$^{2}$) do not show this type of hydrodynamical behaviour. The
values of $\langle p_{T}\rangle$ and $\langle p_{T}^{2}\rangle$
\cite{Jpsi} for the Drell--Yan pairs are smaller than those for $J/\psi$
and $\psi^{\prime}$ and do not change significantly with $E_{T}$.

\vspace{0.3cm} The kinetic freeze--out parameters of pions were extracted
from the analysis of single-- and two--pion spectra measured for central
Pb+Pb collisions at 158 A$\cdot$GeV by the NA49 Collaboration
\cite{ka,hbt,Wi:99}. The results are: $T_{f} \cong 120$ MeV and
$\overline{v}_T \cong 0.55$, i.e. they are very different than those
obtained here from the analysis of heavy hadron spectra. In fact 
we checked that the
parameters obtained for pions lead to the $m_T$ spectra of heavy hadrons
which strongly disagree with the data. A decrease of temperature and an
increase of transverse flow velocity with time is a general property of
expanding systems. The different kinetic freeze--out times of heavy
hadrons and pions allow us to follow the expansion history of the hadron
gas phase created in Pb+Pb collisions at 158 A$\cdot$GeV. The
freeze--out points of heavy hadrons and pions extracted from the data are
plotted in Fig. 2 defining the path of the expanding hadron system in the
$T-\overline{v}_T$ plane.

\vspace{0.3cm} In conclusion, the $m_T$--spectra of $\Omega$, $J/\psi$
and $\psi'$ produced in Pb+Pb collisions at 158 A$\cdot$GeV are analysed
within the hydrodynamical model of the QGP expansion and hadronization.
The spectra are in agreement with the hypothesis of kinetic freeze--out of
these heavy hadrons occurring directly after the transition from the quark
gluon plasma to the hadron gas. A mean collective transverse flow of
hadronizing matter of $\overline{v}_T \cong 0.2$ is extracted from the
fit to the spectra using temperature $T_H \cong$170~MeV fixed by the
analysis of hadron multiplicities \cite{HG}. This result together with a
previously obtained parameters of pion freeze--out ($T_{f} \cong 120$~MeV
and $\overline{v}_T\cong 0.55$) allow for the first time to establish the
history of the expanding hadron matter created in nuclear collisions.

In the RHIC energy range the temperature parameter is approximately the
same $T=T_H\cong 170$~MeV \cite{HG1}, whereas the transverse hydrodynamic
flow at $T=T_H$ is expected to be stronger. The model predictions for
the $m_T$--spectra of
$\Omega$ and charmonia  at the RHIC energies
will be presented elsewhere.

\vspace{0.3cm}
{\bf Acknowledgements}

We are thankful to P.~Bordalo, O.~Drapier, R.~Fini, R. Litava and
D.~R\"ohrich for providing
us with the numerical data of NA50 \cite{Jpsi} and WA97 \cite{Omega}.
The authors are thankful to
L.~Bravina, A.~Dumitru, D.H.~Rischke,  H.~Stroebele,  D.~Teaney and Nu Xu
for
useful discussions. M.I.G. is thankful to the Humboldt Foundation for the
financial support.
The research described in this publication was made possible in part by
Award \# UP1-2119 of the U.S. Civilian Research and Development
Foundation for the Independent States of the Former Soviet Union
(CRDF) and INTAS grant 00-00366.

\newpage

\begin{figure}[p]
\epsfig{file=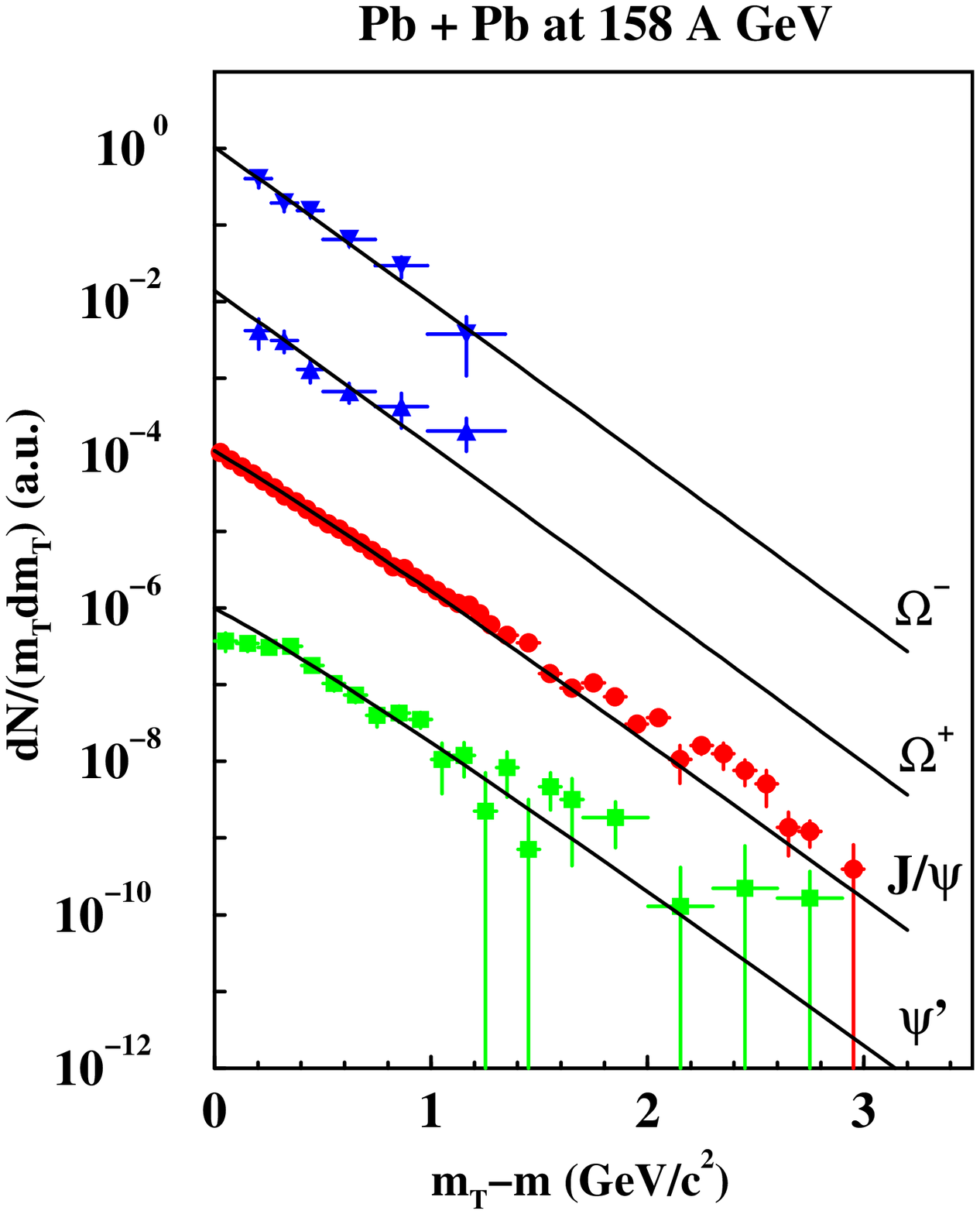,width=18cm} \caption{ The transverse mass spectra of
$\Omega^-$  (triangles down) 
and $\Omega^+$ (triangles up) \protect\cite{Omega} 
 as well as $J/\psi$ (circles)
and $\psi'$ (squares) \protect\cite{Jpsi} produced in Pb+Pb collisions at
158 A$\cdot$GeV. The solid lines indicate a prediction of model
(\ref{hydro1}) assuming kinetic freeze--out of heavy hadrons directly
after hadronization of expanding quark gluon plasma. The freeze--out
parameters are: $T = 170$ MeV and $\overline{v}_T = 0.194$. } \label{fig1}
\end{figure}

\begin{figure}[p]
\epsfig{file=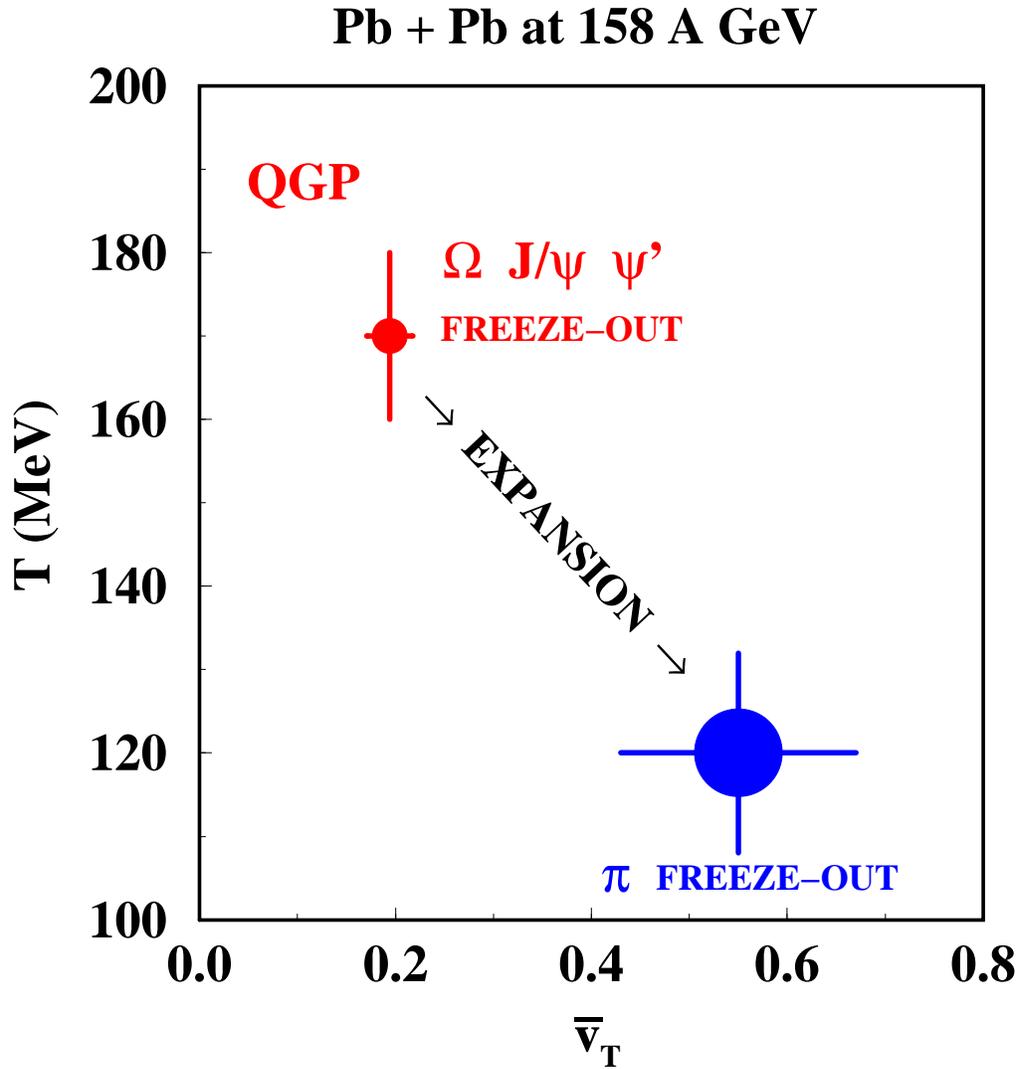,width=18cm} \caption{ The expansion history of
strongly interacting matter created in Pb+Pb collisions at 158
A$\cdot$GeV. The points indicate the temperature $T$ 
and the mean transverse flow
velocity  $\overline{v}_T$ of matter at 
the time of $\Omega$, $J/\psi$ and $\psi'$
freeze--out (upper point) 
and at the time of pion kinetic freeze--out (lower
point). 
} \label{fig2}
\end{figure}

\end{document}